\documentclass[sigconf,screen]{acmart}

\AtBeginDocument{%
  }

\setcopyright{cc}
\setcctype{by}
\acmDOI{10.1145/3843775.3844544}
\acmYear{2026}
\copyrightyear{2026}
\acmISBN{979-8-4007-2987-4/2026/10}
\acmConference[AISM '26]{Proceedings of the 2nd International Workshop on AI for Software Modernization}{October 12--16, 2026}{Munich, Germany}
\acmBooktitle{Proceedings of the 2nd International Workshop on AI for Software Modernization (AISM '26), October 12--16, 2026, Munich, Germany}
\acmSubmissionID{asews26aismmain-p3-p}
\received{2026-08-11}
\received[accepted]{2026-08-23}

\usepackage{tikz}
\usetikzlibrary{positioning, arrows, decorations.pathreplacing, calc}
\usepackage{amsmath}

\usepackage{xspace}
\usepackage{graphicx}
\usepackage{array}        % Better column formatting
\usepackage{booktabs}     % Cleaner, professional-looking rules (optional but great)
\usepackage[acronym]{glossaries}
\usepackage[]{hyphenat}
\usepackage{pgfplots}
\pgfplotsset{compat=1.18}
\usepackage{xcolor}

\glsdisablehyper
\newacronym{ub}{UB}{Undefined Behaviour}
\newacronym{ast}{AST}{Abstract Syntax Tree}
\newacronym{abi}{ABI}{Application Binary Interface}

\newcommand{\tinyskip}{\vspace{3pt}}
\newcommand{\mypar}[1]{\tinyskip\noindent\textbf{#1.}\xspace}

\begin{document}

\title{From C to Idiomatic Rust: A Ship-of-Theseus Agentic Translation}

\author{Vasily A. Sartakov}
\correspondingauthor
\orcid{0000-0003-1894-2621}
\affiliation{%
  \institution{Huawei Technologies Research \& Development (UK) Ltd}
  \city{Cambridge}
  \country{United Kingdom}
}
\email{vasily.sartakov@huawei.com}

\begin{abstract}
C underpins operating systems, embedded platforms, and network infrastructure as its abstractions map directly to machine behaviour.
Its explicit memory model, predictable data representations, and minimal runtime allow compilers to generate fast, deterministic code.
These properties also leave correctness and memory safety entirely to the programmer, making undefined behaviour, pointer misuse, and lifetime errors persistent sources of defects and security vulnerabilities in long-lived C codebases.
Rust eliminates most failure modes through a static ownership and borrowing model that enforces memory safety and aliasing constraints at compile time.
However, mature C systems cannot be translated directly: implicit layout assumptions, aliasing patterns, and undefined behaviour must be reconstructed before safe Rust can be produced.

This paper presents a migration methodology that first generates a semantics-preserving, non-idiomatic Rust baseline and then incrementally rewrites it into idiomatic Rust using agentic AI, validating each step through compilation and behavioural testing.
Applied to iodine, a real-world DNS tunnel, the approach demonstrates that reliable C-to-Rust migration is a structured transformation workflow rather than a single translation step.

\end{abstract}

\keywords{C-to-Rust migration, Rust, agentic translation}

\begin{CCSXML}
<ccs2012>
   <concept>
       <concept_id>10011007.10011074.10011111.10011113</concept_id>
       <concept_desc>Software and its engineering~Software evolution</concept_desc>
       <concept_significance>500</concept_significance>
       </concept>
   <concept>
       <concept_id>10011007.10011074.10011099</concept_id>
       <concept_desc>Software and its engineering~Software verification and validation</concept_desc>
       <concept_significance>300</concept_significance>
       </concept>
   <concept>
       <concept_id>10011007.10011074.10011092.10011782</concept_id>
       <concept_desc>Software and its engineering~Automatic programming</concept_desc>
       <concept_significance>100</concept_significance>
       </concept>
 </ccs2012>
\end{CCSXML}

\ccsdesc[500]{Software and its engineering~Software evolution}
\ccsdesc[300]{Software and its engineering~Software verification and validation}
\ccsdesc[100]{Software and its engineering~Automatic programming}

%%% A "teaser" image appears between the author and affiliation
%%% information and the body of the document, and typically spans the
%%% page.
% \includegraphics[width=\textwidth]{sampleteaser}


\maketitle

%-------------------------------------------------------------------------------
\section{Introduction}
%-------------------------------------------------------------------------------

C remains one of the most influential and widely deployed programming languages.
It provides the implementation foundation for operating systems, embedded systems, networking infrastructure, databases, language runtimes, and many other long-lived software systems.
Since its development at Bell Laboratories in the early 1970s, C has remained a fundamental language for systems programming because of its standardisation, portability, and efficient implementation across a wide range of hardware platforms.
%\cite{ritchie88,Harbison2002,ISO9899}.

C's enduring popularity largely stems from its language abstractions, which closely reflect the underlying machine architecture.
The language employs an explicit memory model, provides straightforward procedural control-flow constructs, and defines data representations that are largely predictable within the constraints of the target platform  \gls{abi}.
Consequently, source-level constructs can be related directly to machine-level operations, enabling efficient implementation and optimisation when execution time, memory consumption, or hardware interaction are important.
Together with a compact language definition, a stable standardisation process, and mature compilers such as GCC and Clang/LLVM, these characteristics make C particularly suitable for systems software and other performance-critical applications that require predictable execution costs \cite{Kernighan1988,Gustedt2019}.

Despite its strengths, C provides relatively few language-level guarantees for memory safety or program correctness.
The language permits unrestricted pointer arithmetic and direct manipulation of memory, placing responsibility for maintaining object lifetimes, array bounds, and pointer validity entirely on the programmer.
As a result, many classes of programming errors are not detected either by the compiler or at runtime.

Common sources of software defects include use-after-free errors, out-of-bounds memory accesses, double-free errors, integer overflow,  null-pointer dereferences, and the use of uninitialised memory.
Such errors usually result in \gls{ub}, meaning that the C standard imposes no requirements on the outcome of program execution.
\gls{ub} enables aggressive compiler optimisation but also complicates debugging, testing, and formal reasoning about program correctness.
In security-critical software, violations of memory safety remain among the most common causes of exploitable vulnerabilities, including arbitrary code execution, privilege escalation, and information disclosure.

These limitations have motivated extensive research on static analysis, dynamic analysis, formal verification, and runtime instrumentation techniques for C programs.
Widely adopted tools, including AddressSanitizer, UndefinedBehaviorSanitizer, Valgrind, \newline and numerous static analysers, aim to detect memory errors and \gls{ub} during development.
More recently, %hardware-assisted protection mechanisms and 
memory-safe programming languages have been proposed to reduce the prevalence of these vulnerabilities whilst preserving the performance characteristics required for systems programming.
%\todo{Sanitizers and hw extension are coslty, alternative is type-safe languages --> transition to rust}

Rust has emerged as a modern systems programming language designed to address many of the memory-safety limitations associated with C whilst retaining comparable performance characteristics.
Unlike C, Rust introduces a strict ownership and borrowing model that enables compile-time verification of memory access patterns, object lifetimes, and data sharing without relying on a garbage collector.
This approach eliminates many common classes of vulnerabilities, including use-after-free errors, double-free errors, and data races, whilst preserving low-level control over memory layout and execution behaviour.

Rust's main advantage is the ability to provide stronger safety guarantees before program execution.
The compiler performs extensive static analysis and rejects programs that violate ownership or borrowing rules, shifting error detection from runtime to compilation.
This design improves reliability for security-critical and performance-sensitive software, particularly in areas traditionally dominated by C, such as operating systems, embedded platforms, and network infrastructure.

Despite advantages, Rust also introduces additional complexity.
The ownership and borrowing model requires programmers to adopt new concepts and can increase initial development effort compared with conventional C programming.
However, the safety guarantees provided by Rust create a strong motivation for migrating parts of existing C infrastructure.
Replacing memory-unsafe components with safe implementations could significantly reduce vulnerabilities whilst maintaining the performance characteristics.
% required for systems software.
Such a transition could be particularly valuable in security-critical domains where long-lived C codebases continue to represent a significant source of memory-related defects.

Nevertheless, migration from C to Rust is not a simple translation problem.
A direct conversion process cannot preserve the behaviour of a mature C codebase whilst simultaneously guaranteeing Rust safety properties.
Existing C programs often depend on implicit assumptions about memory layout, pointer manipulation, object lifetimes, and aliasing patterns that are not explicitly represented in the source code.
In addition, undefined behaviour, manual resource management, compiler-specific extensions, and hardware-dependent optimisations are common in low-level C software and may require substantial redesign when expressed in Rust.
The difference between the two programming models means that successful migration requires analysis and transformation of program structure rather than a purely syntactic conversion.

Recent advances in large language models have demonstrated strong capabilities for code generation and refactoring, making them a promising technology for C-to-Rust migration.
However, most existing approaches formulate migration as direct translation from C to idiomatic Rust, placing both semantic recovery and language transformation into a single step.
For realistic systems software, this objective remains difficult to achieve reliably.

This paper adopts a different perspective.
Rather than translating C directly into idiomatic Rust, we treat migration as the transformation of a semantics-preserving but non-idiomatic Rust program into an idiomatic one.
We define explicit idiomaticity criteria and present a methodology that incrementally replaces non-idiomatic components using agentic AI whilst preserving a continuously executable hybrid codebase.
Each transformation is validated through compilation and behavioural testing before becoming the basis for subsequent migration.

The proposed methodology is implemented as an integrated tool\-chain and evaluated on iodine, a production-quality DNS tunnel comprising approximately 12.5 kSLOC of C code.
The study demonstrates that large-scale migration is better approached as a tooling and verification problem than as a single code-generation task, enabling gradual replacement of legacy C constructs with idiomatic Rust whilst preserving system correctness.

\begin{table*}[t]
\caption{Key differences between C and Rust relevant to automated translation.}
\label{tab:c-rust-diff}
\small
\centering
\setlength{\tabcolsep}{0pt}
\begin{tabular*}{\textwidth}{@{\extracolsep{\fill}} p{0.20\textwidth} p{0.40\textwidth} p{0.40\textwidth} @{}}
\toprule
\textbf{Dimension} & \textbf{C} & \textbf{Rust} \\
\midrule
Memory safety   & No bounds checking, UB on overflow                  & Bounds-checked slices, overflows panic \\
Type safety     & \texttt{void*} erases types, implicit casts           & No \texttt{void*}, no implicit conversions \\
Initialisation  & Uninitialised reads are UB                            & All values initialised before first use \\
Resource mgmt   & Manual \texttt{malloc}/\texttt{free}, leak/double-free & Ownership RAII, compiler-enforced drop \\
Nullability     & Every pointer may be \texttt{NULL}                     & \texttt{Option<T>}, \texttt{\&T} never null \\
String model    & NUL-terminated \texttt{char*}, \texttt{strlen} scans   & \texttt{\&str}/\texttt{\&[u8]}, length in type \\
Global state    & \texttt{static} mutable, no synchronisation           & \texttt{static mut} needs \texttt{unsafe}, \texttt{Mutex} \\
\bottomrule
\end{tabular*}
\end{table*}

%-------------------------------------------------------------------------------

\section{Background}
%-------------------------------------------------------------------------------

C and Rust follow fundamentally different programming paradigms.
C relies on manual memory management and unrestricted low-level access, whilst Rust enforces memory safety and concurrency guarantees through its ownership model.
This division defines the technical and conceptual challenges involved in translating software from one language to another.
Table~\ref{tab:c-rust-diff} summarises the key ones.

\mypar{Memory safety}
C permits arbitrary pointer arithmetic: a function receiving \texttt{char *buf, int~len} can conceptually read or write \texttt{buf[len+1]} or \texttt{buf[-1]}.
Rust's slice type \texttt{\&[u8]} combines~a~pointer with its length, and all accesses are bounds-checked at runtime.
Out-of-bounds access in C results in \gls{ub} and may not trigger a memory protection fault, whereas in Rust it is guaranteed in safe code.
Consequently, a Rust program requires \emph{well-defined bounds}, which are not always recoverable from C.

\mypar{Type safety}
The \texttt{void*} type in C performs type erasure in the sense that C retains no type information at runtime: a function taking \texttt{void *data} operates on untyped memory, and the actual type must be coordinated by convention between caller and callee.
Rust has no analogue to \texttt{void}. 
Its generics use parametric polymorphism, preserving type information for the compiler and enabling static checking (\texttt{fn<T>(x:~\&T)}).
Translating uses of \texttt{void} therefore requires reconstructing the \emph{intended types}, which is a semantic task rather than a syntactic one.

\mypar{Initialisation}
C considers uninitialised memory a valid program state.
Reading an uninitialised variable is \gls{ub}, but the compiler does not prevent it.
Rust, in turn, requires every variable to be initialised before use, and the compiler rejects code paths that might read uninitialised memory.
For translation, it is a \emph{data-flow analysis} problem: it must identify which C variables are definitely initialised before the first use.

\mypar{Resource management}
Rust offers an ownership system that encodes allocation lifetime in the type: every value has exactly one owner, and the compiler inserts deallocation at the end of the owner's scope.
C uses %non-controlled 
manual allocation and deallocation using \texttt{malloc}/\texttt{free} --- a source of \gls{ub} caused by use-after-free or double free.
Translating manual memory to ownership requires inferring the lifetime of each allocation and goes beyond tracing the life-cycle of allocated memory: it must reconstruct how resources flow through pointers, aliases, and calling conventions, and map these behaviours onto Rust's stricter \emph{ownership and borrowing} rules.

\mypar{Nullability}
Every C pointer can be \texttt{NULL}.
A function accepting \texttt{char *s} must check for \texttt{NULL}; otherwise, dereferencing it results in \gls{ub}.
Rust separates nullable from non-nullable references at the type level: \texttt{\&T} is guaranteed non-null; \texttt{Option<\&T>} explicitly represents a value that may be absent.
Translating C pointers requires determining whether \texttt{NULL} is a valid input, which is another semantic challenge.

\mypar{String model}
C represents strings as \texttt{char*} terminated by a NUL byte (\texttt{\textbackslash0}).
String length is computed at runtime via \texttt{str(n)len}, which scans until it finds the terminator or hits the limit.
Rust distinguishes UTF-8 strings (\texttt{\&str}) from arbitrary byte sequences (\texttt{\&[u8]}), with length encoded in the type.
A translator must decide, per buffer, whether it holds a C string (NUL-terminated, possibly UTF-8) or a byte array (length-known, arbitrary content).
Consequently, translating C strings requires inferring the \emph{intended string semantics}, not merely the memory layout.

\mypar{Global state}
C \texttt{static} variables are mutable by default, and the \texttt{const} qualifier is optional and frequently omitted.
Unsynchronised access to a shared state across threads is a data race, and C offers no language-level support to enforce correct synchronisation. 
Idiomatic Rust models shared state explicitly, using \texttt{Atomic*} types for scalar concurrency primitives and \texttt{Mutex<T>} for compound data.
A translator must therefore classify each global according to its effective access pattern: (1) read-only (eligible for \texttt{const}), (2) atomic scalar, or (3) shared mutable state requiring \texttt{Mutex}.
Therefore, mapping of C unconstrained global mutability onto Rust's structured concurrency model requires reconstructing the program's \emph{intended sharing discipline}, rather than translating the declarations. 

These dimensions define the gap that any C-to-Rust translation approach must bridge, and most of them are semantic: they require preserving the intended behaviour of the C code rather than merely transforming its syntax.
As it will be shown in the next section, with proper automation and translation methodology, the gap can be bridged with modern Large Language Models.

\begin{figure*}[t]
\centering
\small
\begin{tabular}{p{0.30\textwidth}p{0.30\textwidth}p{0.30\textwidth}}
\toprule
\textbf{C (original)} & \textbf{c2rust output} & \textbf{Idiomatic Rust} \\
\midrule
\vspace{-1.0em}
\begin{minipage}[t]{0.30\textwidth}
\footnotesize
\begin{verbatim}
int readname(
    char *packet,
    int packetlen,
    char **src,
    char *dst,
    int length)
{
    int len = **src;
    if (len > 63) {
        int ptr = (len & 0x3f)
                  << 8
                  | *(*src + 1);
        return readname(
            packet, packetlen,
            &packet + ptr,
            dst, length);
    }
    memcpy(dst, *src + 1, len);
    *src += len + 1;
    dst[len] = '\0';
    return len;
}
\end{verbatim}
\end{minipage}
&
\vspace{-1.0em}
\begin{minipage}[t]{0.30\textwidth}
\footnotesize
\begin{verbatim}
pub unsafe extern "C" fn
readname(
    packet: *mut c_char,
    packetlen: c_int,
    src: *mut *mut c_char,
    dst: *mut c_char,
    length: c_int)
 -> c_int
{
    let len = **src as c_int;
    if len > 63 {
        let ptr = ...;
        return readname(
            packet, packetlen,
            packet.offset(
                ptr as isize),
            dst, length);
    }
    memcpy(dst as *mut _,
           (*src).offset(1)
               as *const _,
           len as size_t);
    *src = (*src).offset(
        (len + 1) as isize);
    *dst.offset(
        len as isize) = 0;
    len
}
\end{verbatim}
\end{minipage}
&
\vspace{-1.0em}
\begin{minipage}[t]{0.30\textwidth}
\footnotesize
\begin{verbatim}
pub fn readname(
    packet: &[u8],
    src_pos: &mut usize,
    dst: &mut [u8],
    length: usize)
 -> i32
{
    let mut src = *src_pos;
    let c = packet[src];
    src += 1;
    if c & 0xc0 == 0xc0 {
        let offset = ...;
        let mut dummy = offset;
        return readname_loop(
            packet, &mut dummy,
            &mut dst[..],
            length, 10);
    }
    let n = c as usize;
    dst[..n].copy_from_slice(
        &packet[src..src + n]);
    src += n;
    *src_pos = src;
    n as i32
}
\end{verbatim}
\end{minipage}
\\
\bottomrule
\end{tabular}
\caption{Three versions of the same DNS name decoding function. The c2rust
version preserves C's raw pointer semantics and FFI dependencies; the
idiomatic version uses slices, safe indexing, and zero \texttt{unsafe}.}
\Description{Three code listings of the same DNS name-decoding function: the original C, the c2rust output with raw pointers and FFI calls, and the safe idiomatic Rust version using slices and bounds-checked indexing.}
\label{fig:readname}
\end{figure*}

%-------------------------------------------------------------------------------
\subsection{The c2rust Transpiler}
%-------------------------------------------------------------------------------
As will be shown in Section~\ref{method}, the ultimate approach for the translation from C to Rust is, in fact, a transition from non-idiomatic Rust to the idiomatic one.
This is possible because of lossless translation from pure C to non-idiomatic Rust at Clang \gls{ast} level, with further iterative modification of software components on a per-function basis.

The c2rust~\cite{c2rust} transpiler deals with the \gls{ast} of compiling C functions.
It translates the C code into Clang's internal representation, translates the C \gls{ast} into the Rust one, and then emits the Rust code for the corresponding \gls{ast}.
The transpiler preserves the control flow graph, the data flow, and the memory model of the original C:
(1) functions are declared as \texttt{extern "C"} ABI,  (2) global variables become \texttt{static mut}, (3) the pointer arithmetic is possible via \texttt{.offset()}, and (4) the C standard library is called via \texttt{libc}~FFI.

The transpiled code does not introduce any safety.
For example, during the transpiling, the C \texttt{ptr[n]} and \texttt{ptr + n} become Rust \texttt{ptr.offset(n as isize)}.
The pointer arithmetic is preserved, but bounds are not introduced as this information is not available to the transpiler.
As a result, functions become \texttt{unsafe} by default.
Ultimately, the transpiler just generates C code in Rust syntax, but it is a crucial step for the translation.

\subsection{Translation Challenges in Action}

Let us consider a simplified version of a code ported in this project.
It represents typical programming approaches for system software: a buffer manipulation function that decodes a DNS name from wire format.
Figure~\ref{fig:readname} shows the same function in three forms: the original C, the c2rust version, and the final idiomatic Rust after the porting.
They illustrate four patterns that resist mechanical translation:

\mypar{1. Pointer arithmetic as cursor}
The C code uses \texttt{**src} (double indirection) to implement a mutable cursor into a buffer.
The caller owns the buffer whilst the cursor tracks a position within it.
c2rust preserves this as \texttt{*mut *mut c\_char}: an unchecked raw pointer.
The idiomatic version replaces this with \texttt{\&mut usize}, translating the index into a \texttt{\&[u8]} slice.
The translation requires recognising that \texttt{*mut *mut c\_char} is not a generic pointer-to-pointer, but a specific cursor pattern.

\mypar{2. Implicit length through pointer difference}
C code passes \texttt{packet} and \texttt{packetlen} separately and computes \texttt{\&packet+}\allowbreak\texttt{ptr} via pointer arithmetic.
c2rust translates this to \texttt{packet.offset(}\allowbreak\texttt{ptr as isize)}, which is not checked against bounds, and relies on the caller to provide a valid offset.
The idiomatic version fuses the pointer and length into a single \texttt{\&[u8]} slice, eliminating the out-of-bounds class of 
bugs at the type level.
Such an operation requires analyses of the API and propagation of the bounds.

\mypar{3. Manual memory copy via memcpy}
The C code uses \texttt{memcpy} to copy label bytes.
c2rust preserves this as an FFI call to \texttt{libc}, which preserves all risks associated with unsafe memory modification.
The idiomatic version uses \texttt{copy\_from\_slice}, which panics on overflow rather than producing \gls{ub}.

\mypar{4. Null-terminated strings}
The C code writes \texttt{dst[len] = '{\textbackslash}0'} and relies on the caller to provide a large enough buffer.
c2rust translates this to \texttt{*dst.offset(len as isize) = 0} with no checks.
The idiomatic version handles the NUL terminator explicitly within the slice, and callers must account for it.
This snippet also represents the challenges of UTF-8 strings: DNS names are non-UTF-8, thus \texttt{\&[u8]} is used rather than \texttt{\&str}.

These four examples are endemic to C systems software.
Next, we consider our methodology to resolve issues one by one.

\section{Methodology}

We consider the translation of C to idiomatic Rust as a tooling problem of translation from non-idiomatic Rust to the idiomatic one.
Therefore, we introduce criteria to define the idiomatic Rust, and assume that a program is properly implemented if it passes all functionality tests: produces the same output for the same input as the original program.
The tooling component of the system works as a trial-and-error loop where the translating codebase always satisfies the functionality correctness criteria, whilst its components are replaced one-by-one on a per-function basis using LLM-powered reimplementation to satisfy the idiomaticity criteria.

The loop is \emph{agentic}: the LLM system proposes an implementation, compiles it, runs the runtime tests, and revises its own candidate from the feedback (compiler errors or trace divergence) without per-step human action. Human involvement is confined to orchestration: the next migration unit and the architectural policy.

%Below we consider key components and stages, beginning with the definition of idiomatic Rust.

%-------------------------------------------------------------------------------
\subsection{Criteria for Idiomatic Rust}
\label{criteria}
%-------------------------------------------------------------------------------

We define idiomatic Rust along five criteria.
These serve as the reference point for evaluating our methodology and its ability to produce Rust code that aligns with established language practices.

\mypar{C1: Zero \texttt{unsafe} in function signatures}
A function's type signature must not expose \texttt{unsafe} to the caller.
Localised \texttt{unsafe \{...\}} blocks remain acceptable for operations that inherently require them (e.g., FFI calls), but the function must present a safe interface.
This ensures that safety invariants are enforced at the API boundary rather than delegated to the caller.

\begin{figure*}[t]
\centering
\small
\begin{tikzpicture}[
    box/.style={rectangle, draw, thick, rounded corners=2pt,
                align=center, minimum height=0.9cm, font=\footnotesize},
    arrow/.style={->, thick},
    darrow/.style={->, thick, dashed},
]

% Row 1: Main pipeline
\node[box, minimum width=1.8cm, fill=blue!5] (baseline) at (-5.8, 0)
    {Verified\\Baseline};
\node[box, minimum width=1.8cm, fill=blue!5] (leaves) at (-3.2, 0)
    {Leaf-based\\Analysis};
\node[box, minimum width=2.0cm, fill=green!5] (gateA) at (-0.2, 0)
    {Gate A\\Compilation};
\node[box, minimum width=1.6cm, fill=green!5] (gateB) at (2.8, 0)
    {Gate B\\Runtime};
\node[box, minimum width=1.8cm, fill=blue!5] (next) at (5.4, 0)
    {Next Unit\\or Phase};

% Row 1b: Initial translation LLM (above main pipeline)
\node[box, minimum width=2.8cm, fill=cyan!5] (init_llm) at (-1.7, -1.5)
    {Agentic LLM\\\footnotesize describe + implement\\\footnotesize revise + fix};

% Main arrows
\draw[arrow] (baseline) -- (leaves);
\draw[arrow] (leaves) -- (init_llm);
\draw[arrow] (init_llm) -- (gateA);
\draw[arrow] (gateA) -- node[above, font=\tiny] {pass} (gateB);
\draw[arrow] (gateB) -- node[above, font=\tiny] {pass} (next);

% Constraints below gates
%\node[font=\tiny, xshift=6pt, below=2pt of gateA] {enforces C1, C2, C4, C5};

% Row 2: Diagnostic feedback loop (label order 3, 2, 1 left to right)
%\node[box, minimum width=3cm, fill=orange!5, minimum width=1.6cm] (llm) at (-0.2, -2.8) {3. LLM revises};

\node[box, minimum width=3cm, fill=red!5, minimum width=1.6cm] (trace) at (2.8, -2.8)
    {Trace: old vs. new};

%\draw[arrow] (trace) -- (check);
%\draw[arrow] (check) -- (init_llm);

\draw[darrow] (trace.west) to[out=180, in=-80]
    node[font=\tiny, pos=0.45, above] {Divergence} (init_llm.south);

\draw[darrow] (gateA.west) to[out=180, in=90]
    node[font=\tiny, pos=0.45, above] {revise} (init_llm.north);

% Feedback arrows
\draw[darrow] (gateB.south) to[out=-45, in=90]
    node[font=\tiny, pos=0.45, right] {fail} (trace.north);

% Annotation bracket
%\draw[thick, decorate, decoration={brace, mirror}]
%    ($(gateA.south west) + (0, -1.2)$) -- ($(gateC.south east) + (0, -1.2)$)
%    node[midway, below, font=\footnotesize]
%    {Phase 1: C1, C2, C4, C5 $\rightarrow$ Phase 2: C3};

\end{tikzpicture}
\caption{Each transformation passes through two gates: compilation (A) and runtime verification (B). When Gate~B fails, the diagnostic loop traces old and new implementations and feeds the divergence back to the language model for revision.}
\Description{Pipeline diagram of the migration loop: a verified baseline feeds leaf-based analysis, an agentic LLM produces the candidate, Gate A checks compilation and Gate B checks runtime behaviour; a divergence trace feeds failures back to the model for revision.}
\label{fig:gate-cycle}
\end{figure*}
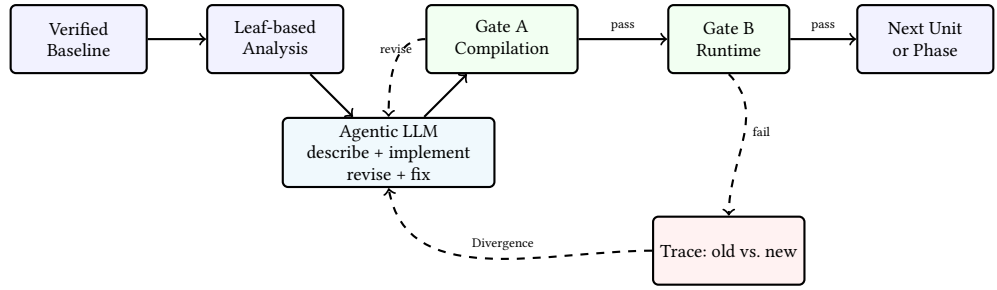

\mypar{C2: Slice-based buffer access}
Buffer parameters must use \texttt{\&[u8]} or \texttt{\&mut [u8]} instead of raw C pointers such as \texttt{const c\_char} or \texttt{mut c\_char}.
Encoding length in the type eliminates the need for manual bounds management and prevents out-of-bounds access at the call site.

\mypar{C3: No \texttt{static mut}}
Global mutable state must not rely on \texttt{static mut}, which is incompatible with Rust's aliasing and concurrency guarantees.
Simple global scalars use atomic types, for example \texttt{AtomicUsize}, whilst compound global state is wrapped in synchronisation primitives such as \texttt{Mutex<T>}.
This removes a major source of \gls{ub} inherited from C.

\mypar{C4: No \texttt{extern "C"} on internal functions}
Only functions that participate directly in the FFI boundary (e.g., callbacks invoked from C) may use the \texttt{extern "C"} ABI.
All internal Rust functions use the Rust ABI, preserving type checking and calling conventions within the Rust portion of the codebase.

\mypar{C5: No C string or memory functions}
C library routines such as \texttt{strlen}, \texttt{strcpy}, \texttt{memcpy}, \texttt{malloc}, and \texttt{free} are replaced with Rust standard library abstractions.
String data is represented as \texttt{\&str} when UTF-8 validity is guaranteed, or as \texttt{\&[u8]} when arbitrary byte sequences are required.
This removes unchecked pointer arithmetic and manual memory management from safe code.

These five criteria are not exhaustive, but they capture the essential properties that distinguish idiomatic Rust from mechanically translated c2rust output.
A codebase satisfying all five is safe by construction in its safe regions, with \texttt{unsafe} confined to the FFI boundary where it is unavoidable.

\subsection{Translation Stages}
\label{method}

The pipeline comprises three components: (1) an automatic transpiler (c2rust) that produces compilable but non-idiomatic Rust from C source, (2) a language model that performs per-function code transformations, and (3) an infrastructure that provides automated gates, runtime tests, and trace-based diagnostics (Figure~\ref{fig:gate-cycle}).

\mypar{Verified baseline}
The functional correctness is the fundamental oracle against which every subsequent transformation is validated.
The building infrastructure is responsible for compiling the project in various configurations: baseline, c2rust version, intermediate step version, versions with instrumentation.
The latter is important to check the \emph{functional equivalence} at the function level: producing the same output and state modification for the same input.

\mypar{Leaf-based ordering}
The transition is performed on per-function basis.
A static analysis of the c2rust codebase produces a directed call graph in which we select functions with no internal dependencies -- leaf nodes.
The leaf functions carry minimal risk of unintended side effects. 
However, the transition to idiomatic Rust impacts the caller functions as arguments passed into leaf functions may require bounds.

Once all leaves in a given dependency tier are ported, the next tier -- functions whose callees now reside entirely in the safe subset -- becomes available.
This bottom-up ordering guarantees that each transformation rests on a foundation of already-verified safe code.

\mypar{Candidate Implementation} We feed the leaf functions one by one into the LLM system.
The system should describe what a particular function does, and then implement this functionality in idiomatic Rust.
The context includes the criteria for idiomaticity, the source code of the C and c2rust versions, and tests if available.
The process is iterative and agentic, bound with Gate~A.

\mypar{A and B gates}
When the candidate function is implemented, the project must compile with zero errors.
Beyond type-level errors such as signature mismatches, missing imports, incorrect API usage, Gate~A enforces the idiomatic and safety criteria (Section~\ref{criteria}) as hard errors.

The system must run under a realistic workload for a sustained period processing real data.
If it fails, both the old and new implementations are instrumented with equivalent trace output, executed on identical inputs, and compared.
The trace divergence is fed back to the language model, which revises the transformation.
This loop repeats until Gate~B passes.

\mypar{Two-phase separation}
Due to the complexity of the translation processes and limits of language models, any complex modification should be split into a set of basic, atomic steps: ultimately, we consider translation as a tooling problem.
We separate the translation into two phases.

Phase~1 targets function bodies and type signatures, addressing
criteria C1, C2, C4, and C5: removing \texttt{unsafe} from function types (C1), replacing raw pointers with slices (C2), eliminating \texttt{extern "C"} on internal functions (C4), and replacing C string functions with Rust standard library equivalents (C5).
This phase is mechanical and high-leverage: it eliminates the majority of unsafe surface area in a project and can proceed without reasoning about shared mutable state or concurrent access.

Phase~2 targets global mutable state, addressing the C3 criteria.
In this phase, the system converts each \texttt{static mut} declaration to an \texttt{Atomic*} or \texttt{Mutex<T>} wrapper.
This phase is challenging because it requires reasoning about lock ordering and potential deadlocks.
Within Phase~2, the system follows a risk-ascending priority: atomic scalars first (no lock nesting, C3 satisfied statically), then counters (snapshot-after-increment semantics), then simple Mutex wrappers on single-path data, and finally complex Mutex wrappers on data shared across multiple
execution paths.
The rationale is that simpler wrappers are mechanically checkable, whilst complex wrappers require full-system lock-order
analysis.
Completing Phase~1 before Phase~2 keeps this analysis tractable: with all function signatures already safe (C1, C2, C4, C5 satisfied), the lock-wrapping problem (C3) becomes self-contained.

%-------------------------------------------------------------------------------
\section{Implementation}
%-------------------------------------------------------------------------------

We apply the methodology to a concrete case: translating the iodine\footnote{\url{https://code.kryo.se/iodine/}} DNS tunnel from C to idiomatic Rust.
Iodine is a real-world systems program of moderate size --- approximately 12.5 kSLOC of C across 17 source files --- large enough to exhibit structural complexity.
It exercises the full range of C idioms that resist mechanical translation: raw pointer arithmetic, static buffers, function pointers stored in structs (via \texttt{encoder\_ops}), global mutable state, and extensive use of C string and memory routines such as \texttt{strlen}, \texttt{strcpy}, and \texttt{memcpy}.
These are precisely the patterns that c2rust preserves verbatim and that our methodology is designed to eliminate.
The system also has a multi-layer client-server architecture: the client encodes and encapsulates traffic into DNS queries, whilst the server decodes and forwards it, with both sides organised around a \texttt{select()}-based event loop.

This section describes the target architecture, the project-specific challenges that emerged during porting and which helped to generalise the methodology.
We also provide quantitative metrics used to track progress of the translation.

%-------------------------------------------------------------------------------
\subsection{Iodine Architecture}
%-------------------------------------------------------------------------------

Iodine is a DNS tunnel that encapsulates IP packets inside DNS queries and responses, allowing IP traffic to traverse networks that only permit DNS.
The system has two components, \texttt{iodined} (server) and \texttt{iodine} (client).
Both endpoints create and configure a \texttt{tun} device: the client captures outgoing IP packets on its local \texttt{tun}, fragments and encodes them into DNS messages, and sends those messages into the DNS resolution path.
The server receives the encoded payloads (typically routed indirectly via recursive resolvers to the authoritative host), decodes and reassembles the IP packets, and injects them into its own \texttt{tun} interface for forwarding.

When the client sends an IP packet, iodine processes it through four stages.
First, the payload is compressed (zlib via \texttt{compress2} and \texttt{uncompress}) to reduce size as DNS labels and names impose strict length limits compared to typical MTU sizes.
Second, the compressed bytes may be encrypted using a Blowfish cipher keyed by a pre-shared password.
Third, the resulting bytes are encoded into a DNS-safe alphabet (one of Base32, Base64, etc.), trading encoding overhead against compatibility with DNS character restrictions.
Fourth, the encoded data is embedded in a DNS query name (e.g., \texttt{\emph{encoded-data}.\emph{topdomain}}) and sent into the DNS resolution path.
The server extracts the payload from incoming queries, reverses the encode/decrypt/decompress pipeline, and writes the reconstructed IP packet to its tunnel device.
The reverse direction embeds downstream data in DNS response records (TXT, MX, SRV, CNAME, A, NULL), completing the bidirectional tunnel.

\mypar{State machine}
Iodine operates in two principal states.
In the \emph{handshake} state, the client and server negotiate connection parameters: authentication, protocol version, encoding codec, DNS query type, tunnel MTU, and optional EDNS0 extensions.
They use message exchanges with timeouts and retries for each stage.
Once the handshake completes, the system transitions to the \emph{tunnel} state, where raw IP traffic flows bidirectionally: upstream from client to server via DNS queries and downstream via DNS responses.
Both directions use a sliding-window fragmentation and reassembly protocol, as individual packets may be split across multiple messages.

%-------------------------------------------------------------------------------
\subsection{Porting Challenges}
%-------------------------------------------------------------------------------

The methodology presented in this work is a human-in-the-loop translation pipeline developed and generalised during the translation of the iodine project: the per-function transformation and verification loop is automated and agentic, whilst a human provides strategic orchestration.
In this section, we describe project-specific challenges that emerged during this process.
Most were orchestration-level tasks --- type deduplication, lock-wrapping, and C-library globals --- handled by humans as exceptions.

\mypar{c2rust duplicate struct definitions}
c2rust translates each C translation unit independently.
Therefore, structs defined in multiple \texttt{.c} files become multiple, distinct Rust types.
C linker matches struct names, whilst Rust treats separately generated structs as incompatible types, which prevents replacing \texttt{extern "C"} declarations with \texttt{use} imports and thus blocks function porting.
In iodine, \texttt{sockaddr\_storage} appeared in six files, \texttt{tun\_user} in two, and \texttt{packet}, \texttt{connection}, and \texttt{query} in three each.
These duplicates produced a cascading dependency freeze: an unportable function blocks its callers, which in turn block their callers.

We resolved this by introducing a single canonical module, removing duplicate definitions, and adding \texttt{use} imports at call sites.
Deduplication must follow dependency order (leaf types first); thus, dependent functions can be ported at the signature level.
This type-level deduplication is a prerequisite to function porting and is not captured by the leaf-based ordering.

\mypar{FFI adapter scaffolding}
During incremental porting, we encountered a recurring intermediate state.
Safe Rust implementations of functionality (for example \texttt{base32::encode}) existed, but existing callers still passed C function pointers through a C struct such as \texttt{encoder\_ops}, which requires signatures like \texttt{*const encoder}.
To bridge this mismatch, we introduced thin \texttt{\_ffi} wrappers that accept raw pointers, convert them to Rust slices, invoke the safe Rust routine, and write results back into C buffers.
Each wrapper was paired with a \texttt{static mut} global holding the function pointer table and often duplicated the original C struct definitions across compilation units. These adapters enable mixed c2rust and Rust execution and let individual functions be ported early.
These temporary solutions were removed at later stages.

\mypar{Static mut elimination and deadlock hazards}
Phase 2 converts \texttt{static mut} declarations to \texttt{Atomic*} or \texttt{Mutex<T>} wrappers.
The prescribed order is atomics first, then mutexes, and each global must be wrapped one at a time with full runtime verification after each change.
In the iodine client, we wrapped 16 \texttt{static mut} globals, applying 13 atomics followed by 3 \texttt{Mutex<T>} wrappers.
An attempt to wrap all three mutexes in a single batch produced a live deadlock: the tunnel established but transfers stalled at zero bytes.
Reverting to one-at-a-time wrapping with runtime tests exposed two silent deadlock patterns that the compiler does not detect.
Both patterns compile and pass static checks but fail only under real traffic.
The practical rule is strict: perform Phase 2 one global at a time and run full runtime tests after each wrapping.

\mypar{Unwrappable static mut from C state}
POSIX \texttt{getopt()} communicates via globals (\texttt{optarg}, \texttt{optind}, \texttt{opterr}, \texttt{optopt}), which appear in Rust as \texttt{extern "C" \{ static mut optarg: *mut c\_char; static mut optind: c\_int; \}}.
These extern statics cannot be converted to \texttt{Atomic} or \texttt{Mutex<T>} because writes originate inside the C library and bypass any Rust-side wrapper. 
Demoting \texttt{static mut} to immutable \texttt{static} is ineffective since extern static access remains \texttt{unsafe}.
The only sound solution is to remove the C global state by replacing \texttt{getopt()} with a Rust argument parser (for example, iterating \texttt{std::env::args()} and parsing flags).
This is a structural refactor for both client and server entry points.
We deferred it because the extern statics are confined to \texttt{main()}, used single-threadedly, and present a small, auditable \texttt{unsafe} surface. 
In the future, this should affect the methodology to treat C library globals as replacement tasks, not Phase~2 wrapping candidates.

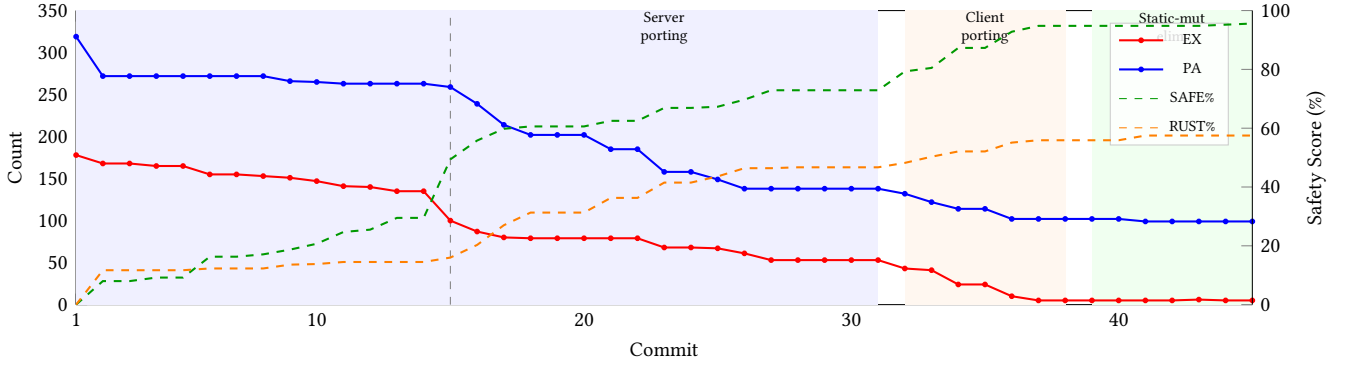
\begin{figure*}
\resizebox{\textwidth}{!}{
\begin{tikzpicture}

% LEFT AXIS: counts (0-350)
\begin{axis}[
    width=16cm,
    height=4.0cm,
    scale only axis,
    xlabel={Commit},
    ylabel={Count},
    ylabel near ticks,
    ymin=0, ymax=350,
    xmin=1, xmax=45,
    xtick={1,10,20,30,40},
    ytick={0,50,100,150,200,250,300,350},
    every axis/.append style={font=\small},
    legend style={at={(0.98,0.75)},anchor=east,font=\scriptsize,
                  draw=gray!40,fill=white,fill opacity=0.85,
                  text opacity=1,row sep=2pt},
    axis background/.style={fill=none},
    axis y line*=left,
]

% phase bands
\fill[blue!6]  (axis cs:1,0)  rectangle (axis cs:31,350);
\fill[orange!6] (axis cs:32,0) rectangle (axis cs:38,350);
\fill[green!6] (axis cs:39,0) rectangle (axis cs:45,350);

% type-level refactor marker (Codec enum + FFI adapter removal)
\draw[gray,dashed] (axis cs:15,0) -- (axis cs:15,350);

%% EX
\addplot[thick,red,mark=*,mark size=0.7] coordinates {
    (1,178)     (2,168)     (3,168)     (4,165)     (5,165)     (6,155)     (7,155)
    (8,153)     (9,151)     (10,147)     (11,141)     (12,140)     (13,135)     (14,135)
    (15,100)     (16,87)     (17,80)     (18,79)     (19,79)     (20,79)     (21,79)
    (22,79)     (23,68)     (24,68)     (25,67)     (26,61)     (27,53)     (28,53)
    (29,53)     (30,53)     (31,53)     (32,43)     (33,41)     (34,24)     (35,24)
    (36,10)     (37,5)     (38,5)     (39,5)     (40,5)     (41,5)     (42,5)
    (43,6)     (44,5)     (45,5)     (46,5)     (47,5)     (48,5)     (49,5)
    (50,5)     (51,5)     (52,5)     (53,5)     (54,5)     (55,5)     (56,5)
    (57,5)     (58,5) 
};
\addlegendentry{EX}

%% PA
\addplot[thick,blue,mark=*,mark size=0.7] coordinates {
    (1,319)     (2,272)     (3,272)     (4,272)     (5,272)     (6,272)     (7,272)
    (8,272)     (9,266)     (10,265)     (11,263)     (12,263)     (13,263)     (14,263)
    (15,259)     (16,239)     (17,214)     (18,202)     (19,202)     (20,202)     (21,185)
    (22,185)     (23,158)     (24,158)     (25,149)     (26,138)     (27,138)     (28,138)
    (29,138)     (30,138)     (31,138)     (32,132)     (33,122)     (34,114)     (35,114)
    (36,102)     (37,102)     (38,102)     (39,102)     (40,102)     (41,99)     (42,99)
    (43,99)     (44,99)     (45,99)     (46,99)     (47,99)     (48,99)     (49,99)
    (50,99)     (51,99)     (52,99)     (53,99)     (54,99)     (55,99)     (56,99)
    (57,99)     (58,99) 
};
\addlegendentry{PA}

%% fake legend entries for the series plotted on the right axis
\addlegendimage{green!60!black,dashed}
\addlegendentry{SAFE\%}
\addlegendimage{orange,dashed}
\addlegendentry{RUST\%}

% phase labels
\node[font=\scriptsize,align=center] at (axis cs:23,330) {Server\\porting};
\node[font=\scriptsize,align=center] at (axis cs:35,330) {Client\\porting};
\node[font=\scriptsize,align=center] at (axis cs:42,330) {Static-mut\\elim.};

\end{axis}

% RIGHT AXIS: percentages (0-100) — raw values, no scaling
\begin{axis}[
    width=16cm,
    height=4.0cm,
    scale only axis,
    xmin=1, xmax=45,
    ymin=0, ymax=100,
    axis y line*=right,
    axis x line=none,
    ylabel={Safety Score (\%)},
    ylabel near ticks,
    ytick={0,20,40,60,80,100},
    every axis/.append style={font=\small},
]

%% SAFE% (coeff 1: (EX+NM)/(EX₀+NM₀), raw % on right axis)
\addplot[thick,green!60!black,dashed] coordinates {
    (1,0.0)     (2,8.0)     (3,8.0)     (4,9.2)     (5,9.2)     (6,16.3)     (7,16.3)
    (8,17.1)     (9,18.7)     (10,20.7)     (11,24.7)     (12,25.5)     (13,29.5)     (14,29.5)
    (15,49.4)     (16,55.8)     (17,59.8)     (18,60.6)     (19,60.6)     (20,60.6)     (21,62.5)
    (22,62.5)     (23,66.9)     (24,66.9)     (25,67.3)     (26,69.7)     (27,72.9)     (28,72.9)
    (29,72.9)     (30,72.9)     (31,72.9)     (32,79.3)     (33,80.5)     (34,87.3)     (35,87.3)
    (36,92.8)     (37,94.8)     (38,94.8)     (39,94.8)     (40,94.8)     (41,94.8)     (42,94.8)
    (43,94.8)     (44,95.2)     (45,95.6)     (46,95.6)     (47,95.6)     (48,95.6)     (49,95.6)
    (50,95.6)     (51,95.6)     (52,95.6)     (53,95.6)     (54,95.6)     (55,95.6)     (56,95.6)
    (57,95.6)     (58,95.6) 
};
%\addlegendentry{SAFE\%}

%% RUST% ((PA+DU)/(PA₀+DU₀), raw % on right axis)
\addplot[thick,orange,dashed] coordinates {
    (1,0.0)     (2,11.7)     (3,11.7)     (4,11.7)     (5,11.7)     (6,12.3)     (7,12.3)
    (8,12.3)     (9,13.6)     (10,13.8)     (11,14.5)     (12,14.5)     (13,14.5)     (14,14.5)
    (15,16.0)     (16,20.3)     (17,27.0)     (18,31.3)     (19,31.3)     (20,31.3)     (21,36.3)
    (22,36.3)     (23,41.5)     (24,41.5)     (25,43.6)     (26,46.4)     (27,46.4)     (28,46.7)
    (29,46.7)     (30,46.7)     (31,46.7)     (32,48.2)     (33,50.3)     (34,52.1)     (35,52.1)
    (36,55.1)     (37,55.9)     (38,55.9)     (39,55.9)     (40,55.9)     (41,57.5)     (42,57.5)
    (43,57.5)     (44,57.5)     (45,57.5)     (46,57.5)     (47,57.5)     (48,57.5)     (49,57.5)
    (50,57.2)     (51,57.2)     (52,57.2)     (53,57.2)     (54,57.2)     (55,57.2)     (56,57.2)
    (57,57.2)     (58,57.2) 
};
%\addlegendentry{RUST\%}

\end{axis}
\end{tikzpicture}
}
\caption{Evolution of the \texttt{SAFE\%} and \texttt{RUST\%} metrics over the commit history.}
\Description{Line chart over 45 commits showing the counts of unsafe constructs decreasing whilst the SAFE-percent and RUST-percent metrics rise through the server-porting, client-porting, and static-mut elimination phases.}
\label{fig:progress}
\end{figure*}

\subsection{Evolution of Safety}
\label{sec:metrics}

We track translation progress over the c2rust baseline using two measures, each defined as the share of initial C-origin artefacts eliminated.
The first, \texttt{SAFE\%}, removes the C ABI surface: \texttt{unsafe extern "C" fn} declarations (\texttt{EX}) and \texttt{\#[no\_mangle]} annotations (\texttt{NM}).
The second, \texttt{RUST\%}, removes C-style unsafe patterns in function bodies: raw pointer arithmetic (\texttt{PA}) and deep unsafe operations such as \texttt{copy\_nonoverlapping} and \texttt{write\_bytes} (\texttt{DU}).

Subscript 0 denotes the initial count in the c2rust baseline, and both measures report percentages of that count:

\begin{equation*}
\begin{aligned}
\mathrm{SAFE\%}=100\Bigl(1-\frac{\mathrm{EX}+\mathrm{NM}}{\mathrm{EX}_{0}+\mathrm{NM}_{0}}\Bigr),  \hspace{0.15cm}
\mathrm{RUST\%}=100\Bigl(1-\frac{\mathrm{PA}+\mathrm{DU}}{\mathrm{PA}_{0}+\mathrm{DU}_{0}}\Bigr)
\end{aligned}
\end{equation*}

We omit the raw \texttt{unsafe} count: porting moves \texttt{unsafe} from function declarations into narrow blocks around FFI calls, so counting it would penalise correct refactoring.

Figure~\ref{fig:progress} plots the two measures per commit (each commit ports one to ten functions, always from the same functional component).
The c2rust baseline holds 178 \texttt{EX}, 73 \texttt{NM}, 319 \texttt{PA}, and 144 \texttt{DU}.
Server porting (commits 1--31) lifts \texttt{SAFE\%} to 72\% and \texttt{RUST\%} to 47\%; client porting (commits 32--41) completes all 168 signatures at \texttt{SAFE\%}=95\%; static-mut elimination (commits 39--45, extending to the end of history) wraps the 37 \texttt{static mut} globals (16 client, 21 server) in atomics or mutexes, lifting \texttt{RUST\%} to its final 57\%.

The steep rise at commit~15 is a type-level refactor --- the \texttt{Codec} enum replacing four duplicated C function-pointer structs and eight FFI adapter wrappers --- rather than incremental function porting.
The plateaus are structural: the residual \texttt{SAFE\%} gap is the five signal-handler \texttt{EX} type aliases and six \texttt{\#[no\_mangle]} entry-point annotations, and the un-recovered \texttt{RUST\%} is 99 \texttt{PA} and 99 \texttt{DU} uses confined to the FFI boundary adapters (tun-device and socket I/O; roughly 75 of the \texttt{PA} sites in \texttt{iodined.rs}), where raw pointers are inherent to interacting with C: indexing the C-allocated per-user struct array via \texttt{USERS.offset(i)} and skipping bytes in DNS wire buffers.

The final codebase is approximately 10.3 kSLOC of idiomatic Rust across 16 files, translated from an original 12.5 kSLOC C codebase, with no \texttt{unsafe extern "C"} function declarations and no \texttt{static mut} left in the evaluated code.
The translation took 37 developer-hours over four working days --- 4.5 functions and roughly 278 lines per hour.

\mypar{Validation} Functional equivalence was verified end-to-end: after each of the 168 function ports and each of the 37 \texttt{static mut} wrappings (16 client, 21 server), Gate~B ran the tunnel against the public resolver under real traffic for 30\,s or more, verified \texttt{curl} download and \texttt{netperf} through the tunnel, and passed a wire-format integration test.
This oracle caught what static checks missed: a batch mutex wrap that compiled but deadlocked under load (two silent lock-pattern defects), a NUL-termination bug in \texttt{send\_query}, a codec-command mismatch in fragsize negotiation, and timeouts on non-UTF-8 hostnames --- each diagnosed from runtime divergence and revised within the loop, reverting to the last verified state where needed.
Client porting and static-mut elimination cut the client's \texttt{unsafe} blocks from roughly 86 to 40.

\mypar{Models} The proposed methodology does not require long context, because it is based on iterative re-initialisation of the context for each new function.
For each function identified at the current leaf level, we create new context filling the precise and short description of the task.
We performed translation using subscription-based plans and the OpenCode-based infrastructure, enhanced by the \texttt{rtk} and \texttt{OAC}.
That a particular model choice does not materially affect the outcome is itself evidence of the loop's autonomy: because Gates~A and~B drive revision from compiler and runtime feedback, the migration is steered by the loop rather than by the model; the best result (fewest iterations) was achieved with Qwen3.6-27B.

%-------------------------------------------------------------------------------
\section{Related Work}
%-------------------------------------------------------------------------------

\mypar{Rule-based C-to-Rust translation}
Compiler-based approaches translate C to Rust using deterministic analyses and transformations.
c2rust~\cite{c2rust} provides the canonical transpilation pipeline by preserving the original program structure and generating functionally equivalent Rust.
Subsequent work extends this through ownership inference~\cite{zhang2023ownership}, alias analysis~\cite{emre2023aliasing}, and specialised transformations for challenging C constructs including output parameters, unions, and C I/O APIs~\cite{hong2024tag,hong2024don,hong2025forcrat}.
These techniques improve the safety and idiomaticity of generated code by addressing specific language features whilst preserving program behaviour.
Nevertheless, these remain compiler-driven, isolated transformations targeting specific translation problems rather than end-to-end migration.
We use c2rust as a lossless translation stage, producing a functionally equivalent but non-idiomatic Rust baseline for refinement.

\mypar{LLM-based C-to-Rust translation}
Recent work employs large language models to generate idiomatic Rust directly from C source code.
Existing systems improve translation quality through iterative generation, compilation feedback, repair loops, and validation~\cite{yang2024exploring,eniser2025translatingrealworldcodellms,shiraishi2024smartc2rust,zhou2025c2rusttv}.
Compared with compiler-based approaches, these methods often produce more natural Rust by leveraging semantic reasoning.
Nevertheless, they primarily frame migration as a code generation problem, with the model synthesising a correct Rust implementation from the original C program.
In contrast, we frame migration as a verified refactoring problem: starting from a mechanically translated Rust, we incrementally transform individual functions into idiomatic Rust whilst preserving behavioural equivalence through compilation and runtime verification.

\mypar{Project-level migration}
Recent research extends C-to-Rust translation beyond individual functions by incorporating repository context, dependency information, build systems, and incremental compilation.
RustMap~\cite{cai2025rustmap}, repository-level translation frameworks~\cite{dehghan2025translating}, and EvoC2Rust~\cite{wang2025evoc2rust} demonstrate that exploiting project-level information substantially improves compilation success and scalability for large codebases.
Some of these approaches further suggest that repository evolution, such as commit history, may provide additional semantic context during translation.
Our work is complementary to these efforts.
Rather than focusing on enriching the translation context, we study the migration workflow itself: dependency-aware function ordering, explicit idiomaticity criteria, incremental replacement of non-idiomatic components, and verification gates that ensure the translated system remains continuously executable throughout the migration process.
In our case study, repository history was not required to perform the migration, suggesting that continuous verification and dependency-aware orchestration can be sufficient without historical development context, but conceptually can be combined.

\section{Conclusion}

C remains the foundation of much deployed systems software, but its unchecked pointer arithmetic, manual memory management, and unconstrained global state make long-term maintenance and security difficult.
Rust offers stronger safety guarantees, yet migrating existing C codebases is hard because idiomatic Rust requires semantic information that C does not encode.

In this paper, we present a translation pipeline that treats C to Rust migration as a sequence of correctness-preserving, per-function replacements.
Starting from a lossless c2rust baseline, we use an agentic loop that rewrites functions into idiomatic Rust whilst enforcing behavioural equivalence through runtime gates.

\section{Future Work}

This work will be extended in four directions: generalisation, veri\hyp{}fication-driven translation, whole-project reasoning, and migration orchestration.
First, we will generalise the methodology beyond the iodine case study.
Iodine is a single, moderate-size systems program, so we will apply the loop to a broad range of C projects spanning different sizes, domains, and concurrency profiles (device drivers and OS components).
Demonstrating transfer across these projects will establish generalisability.
Second, compilation and behavioural testing validate each step, but the process is largely reactive.
Combining runtime testing with static analysis, symbolic execution, or automated test generation could provide stronger correctness guarantees and reduce manual oversight required.

Third, although many functions can be migrated independently, architectural changes --- such as redesigning shared state or replacing C-specific libraries --- require reasoning across the entire codebase. Repository-level information, including dependency graphs and commit history, may help recover developer intent and guide these broader transformations. This information was not needed in our study, but may be important for full-project migration.
Fourth, the orchestration itself must scale: from a single human steering the loop to systems that plan, verify, and coordinate large numbers of incremental transformations whilst keeping the software continuously functional.

\section*{Acknowledgements}

ChatGPT and Copilot were used to proofread sections of this work.

\bibliographystyle{ACM-Reference-Format}
\bibliography{lib}

\end{document}